\documentclass[twocolumn,english,showpacs,floatfix]{revtex4}

\usepackage{babel,amsmath,amssymb}
\usepackage[dvips]{graphics}

\begin{document}

\title{Field propagation in de Sitter black holes}

\author{C. Molina}
\email{cmolina@fma.if.usp.br}

\author{D. Giugno}
\email{dgiugno@fma.if.usp.br}

\author{E. Abdalla}
\email{eabdalla@fma.if.usp.br}
\affiliation{Instituto de F\'{\i}sica, Universidade de S\~{a}o Paulo,
C.P. 66318, 05315-970, S\~{a}o Paulo-SP, Brazil}

\author{A. Saa}
\email{asaa@ime.unicamp.br}
\affiliation{Departamento de Matem\'atica Aplicada, UNICAMP,
C.P.6065, 13083-859, Campinas-SP, Brazil}


\begin{abstract}
We present an exhaustive analysis of scalar, electromagnetic and
gravitational perturbations in the background of Schwarzchild--de
Sitter and Reissner--Nordstr\"om--de Sitter spacetimes. 
The field propagation is considered by means of a
semi--analytical (WKB) approach and two numerical schemes: the
characteristic and general initial value integrations. The results are
compared  near the extreme cosmological constant regime, where 
analytical results are presented. A unifying picture is
established  for the dynamics of different spin fields.
\end{abstract}

\maketitle

\section{Introduction}

Wave  propagation around nontrivial solutions of Einstein equations, 
black holes in particular, is an active field of research (see
\cite{Regge-57,Chandrasekhar,Kokkotas-99} and references therein). 
The perspective of gravitational wave detection in the near future and
the great development of numerical general relativity 
have increased even further the activity on this field. Gravitational
waves should be especially strong when emitted by black holes. The
study of the propagation of perturbations around them is, hence,
essential to provide templates for  gravitational wave identification.
On the other hand, recent astrophysical observations indicate that the
universe is undergoing an accelerated expansion phase, suggesting the
existence of a small positive cosmological constant and that de Sitter
(dS) geometry provides a good description of very large scales of the
universe \cite{accel-exp}. We notice also that string theory has
recently motivated many works on asymptotically anti--de Sitter
spacetimes (see, for instance,
\cite{Cardoso-01,AdS-4d,AdS-other,Wang-01}).       

In this work, we perform an exhaustive investigation
of scalar, electromagnetic, and gravitational
perturbations in the background of Schwarzchild--de Sitter (SdS)
and Reissner--Nordstr\"om--de Sitter (RNdS) spacetimes.
Contrasting with the noncharged case, in the RNdS one
the electromagnetic and gravitational perturbations are
necessarily coupled.
We scan the full range of the
cosmological constant, from the asymptotic flat case
($\Lambda=0$) up to the critical value of $\Lambda$
which characterizes,
 for the noncharged case,
 the Nariai solution \cite{Nariai}.
Two different numerical methods and a higher order WKB analysis are
used. The results are compared near the extreme $\Lambda$ regime,
where analytical results can be obtained.

We recall that for any perturbation in the spacetimes we consider that,
after the initial transient phase, there are two main
contributions to the resulting asymptotic wave \cite{Leaver-86,Ching-95}: 
initially the so-called quasinormal modes, which are suppressed at later
time by the tails.
The first can be understood as candidates for
normal modes which, however, decay (their energy eigenvalues becomes
complex), as in the ingenious mechanism first described by Gamow in
the context of nuclear physics \cite{Gamow-28}. After the initial
transient phase, the properties of the resulting waves
are more related to the background spacetime rather than to the source itself. 

It is well known that for
asymptotically flat backgrounds the tails decay
according to a power law, whereas in a space with a positive cosmological
constant the decay is exponential. Curiously, 
$\ell=0$ modes for scalar fields in dS spacetimes, contrasting
with the asymptotic flat cases, exponentially  approach 
a nonvanishing asymptotic value \cite{Brady-97,Brady-99}. 
We detected, by using a noncharacteristic
numerical integration scheme, a dependence of this asymptotic value
on the initial velocities. In particular, it vanishes for static
initial conditions. Our results are in perfect agreement with
the analytical predictions of \cite{Brady-99}.

The semianalytical analyses of this work were performed by using 
the higher order WKB method proposed by Schutz and Will
\cite{Schutz-85}, and improved by Iyer and Will \cite{Iyer-87,WKB-ap}.
It provides a very
accurate and systematic way to study black hole quasinormal modes. 
We apply it to the study of various perturbation fields in the
nonasymptotically flat dS geometry. Quasinormal modes are 
also calculated according to this approximation, and  the results 
are compared to the numerical ones whenever
appropriate, providing a quite complete picture of the question of
quasinormal perturbations for dS black holes.

Concerning the charged case, we analyze in detail the wave propagation of
the massless scalar field and coupled electromagnetic and gravitational fields
in the RNdS spacetime. An important difference concerning
the dynamics of the electromagnetic and gravitational fields is that there
are no pure modes, since both are interrelated. We will show that the
direct picture of the evolution presents us with perfect agreement
of quasinormal frequencies with those obtained by using the
approximation method suggested in \cite{Schutz-85,Iyer-87}. One
important point assessed is the dependence of the fields' decay on the
electric charge of the black hole, including the asymptotically
flat limit, for which we expected to find traces of a power law tail
appearing between the quasinormal modes and the exponential tail.

Two very recent works overlap our analysis presented here.
A similar WKB approach, presented in 
\cite{Konoplya-03}, 
 was used very recently by
Zhidenko \cite{Zhidenko} to study SdS black holes, giving
results in agreement with ours.
Yoshida and Futamase \cite{Yoshida} used a continued fraction
numerical code to calculate quasinormal mode frequencies, with special
emphasis on high order modes. Our results are also compatible.
Finally, we notice that solutions of the wave equation in a
nontrivial background have also been used to infer intrinsic
properties of the spacetime \cite{qnmbh}. 

The paper is organized as follows. Sec. II provides theoretical
considerations and reviews some well-known results that
were useful to our work; Sec. III briefly explains the numerical
and semianalytical methods employed, followed by Sec. IV, which
presents, in detail, our results on field dynamics for near extreme 
SdS and RNdS
geometries. Sec. V deals with the so-called intermediary region,
where the geometries are not extreme. Data on the SdS limit and on
exponential tails are also presented. Sec. VI deals with the
near asymptotically flat region, and Sec. VII presents
our conclusions.

\section{Metric, fields, and effective potentials}

The metric describing a charged, asymptotically de Sitter spherical black
hole, written in spherical coordinates, is given by
\begin{equation}
\label{metric}
ds^{2}=-h(r)dt^{2}+h(r)^{-1}dr^{2}+r^{2}(d\theta^{2}+\sin^{2}\theta
d\phi^{2}) \ ,
\end{equation}
where the function $h(r)$ is
\begin{equation}
h(r)=1 - \frac{2m}{r} + \frac{q^{2}}{r^{2}} - \frac{\Lambda r^{2}}{3} \ .
\end{equation}
The integration constants $m$ and $q$ are the black hole mass and electric
charge, respectively. If the cosmological constant is positive, we have the
Reissner--Nordstr\"{o}m--de Sitter metric. In this case, $\Lambda$ is usually
written as $\Lambda=3/a^{2}$, where the constant $a$ is the
``cosmological radius.''

The spacetime causal structure depends strongly on the zeros of $h(r)$.
Depending on the parameters $m$, $q$, and $a$, the function $h(r)$ may have
three, two, or even no real positive zeros. For the RNdS cases we are
interested in, 
$h(r)$ has three simple real, positive roots ($r_{c}$, $r_{+}$, and $%
r_{-}$), and a real and negative root
$r_{n}=-(r_{-}+r_{+}+r_{c})$. The horizons $r_{-}$, $r_{+}$, and
$r_{c}$, with $r_{-}<r_{+}<r_{c}$, are denoted Cauchy, event, and
cosmological horizons, respectively. 

For the SdS case  ($q=0$), and assuming $m > 0$ and $0 < 9 m^{2}
\Lambda < 1$, the function $h(r)$ has two positive zeros $r_{+}$ and
$r_{c}$ and a negative zero $r_{n} = -(r_{+} + r_{c})$. This is the SdS
geometry in which we are interested.  The horizons $r_{+}$ and
$r_{c}$, with $r_{+}<r_{c}$, are denoted the event and cosmological
horizons, respectively. In this case, the constants $m$ and $a$ are
related to the roots by  
\begin{equation}
a^2 = r_{+}^2 + r_{c}^2 + r_{+} r_{c}\ ,
\end{equation}
\begin{equation}
2 m a^2 = r_{+} r_{c} (r_{+} + r_{c})\ .
\end{equation}
If $9 m^{2} \Lambda = 1$, the zeros $r_{+}$ and $r_{c}$ degenerate into a
double root. This is the extreme SdS black hole. If $m^2\Lambda > 1$,
there are no real positive zeros, and the metric (\ref{metric}) does
not describe a black hole. 

In both the SdS and RNdS cases,  we shall study
the perturbation fields in the exterior region, defined as
\begin{equation}
T_{+} = \left\{ (t,r,\theta,\phi),r_{+}<r<r_{c}\right\} \ .
\end{equation}
In this region $T_{+}$, we define a ``tortoise coordinate'' $x(r)$ in the
usual way,
\begin{eqnarray}
x(r) & = & - \frac{1}{2\kappa_{c}} \ln(r_{c} - r) + \frac{1}{2\kappa_{+}}
\ln(r - r_{+})  \nonumber \\
& & \mbox{} - \frac{1}{2\kappa_{-}} \ln(r - r_{-}) + \frac{1}{2\kappa_{n}}
\ln(r - r_{n}) \ ,
\end{eqnarray}
with
\begin{equation}
\kappa_{i} = \frac{1}{2} \left| \frac{d h(r)}{dr}\right|_{r=r_{i}} \ .
\end{equation}
The constants $\kappa_{-}$, $\kappa_{+}$, and $\kappa_{c}$ are the surface
gravities associated with the Cauchy, event, and cosmological horizons,
respectively. For the SdS case, the term associated with the Cauchy
horizon is absent.

Consider now a scalar perturbation field 
$\Phi$ obeying the massless Klein-Gordon equation 
\begin{equation}
\Box\Phi = 0 \ . \label{boxphiequalzero}
\end{equation}
The usual separation of variables in terms of a radial field and a
spherical harmonic $\textrm{Y}_{\ell,m}(\theta,\varphi)$,
\begin{equation}
\Phi=\sum_{\ell\,
m} \frac{1}{r} \psi_{\ell}^{sc}(t,r)\textrm{Y}_{\ell}(\theta,\phi)  \ ,
\label{Ansazs_field}
\end{equation}
leads to 
Schr\"{o}dinger-type equations in the tortoise coordinate
for each value of $\ell$,
\begin{equation}
-\frac{\partial^{2}\psi_{\ell}^{sc}}{\partial
 t^{2}}+\frac{\partial^{2}\psi_{\ell}^{sc}}{\partial
 x^{2}}=V^{sc}(x)\psi_{\ell} \ ,
\label{wave_equation}
\end{equation}
where the effective potential $V^{sc}$ is given by
\begin{equation}
V^{sc}(r)=h(r)\left[\frac{\ell(\ell + 1)}{r^2}+\frac{2m}{r^{3}} - \frac{2q^2%
}{r^{4}} - \frac{2}{a^{2}}\right] \ .
\end{equation}

The situation for higher spin perturbations is quite different.
In the SdS geometry, in contrast to the case
of an electrically charged black hole,  it is possible to have pure
electromagnetic and gravitational perturbations. 
For both cases, we have Schr\"{o}dinger-type effective equations.
For the first, the
effective potential is given by \cite{Ruffini-72} 
\begin{equation}
V^{el}(r) = h(r)\frac{\ell(\ell + 1)}{r^{2}},
\end{equation}
with $\ell\ge 1$.
The gravitational 
perturbation theory for the exterior  Schwarzschild--de Sitter  geometry
has been developed in \cite{Chandrasekhar,Cardoso-01}. The potentials
for the axial and polar modes are, respectively,
\begin{equation}
V^{ax}(r) =
h(r)\left[\frac{\ell(\ell+1)}{r^{2}}-\frac{6m}{r^{3}}\right] \ , 
\end{equation}
\begin{eqnarray}
V^{po}(r) & = & \frac{2 h(r)}{r^{3} \left(3m+cr\right)^{2}} 
\left[ 9m^{3}+3c^{2}mr^{2}+c^{2}(1+c)r^{3} \right. \nonumber \\
& & \mbox{} \left. +3m^{2} (3cr-\Lambda r^{3}) \right]  \ ,
\end{eqnarray}
with $2c = (\ell-1)(\ell+2)$ and $\ell\ge 2$.
For perturbations with $\ell > 0$, we can show explicitly that all the
effective potentials $V(x) \equiv V(r(x))$ are positive definite. For
scalar perturbations with $\ell = 0$, however, the effective potential has one
zero point $x_{0}$  and it is negative for $x > x_{0}$.

The perturbation theory for the RNdS geometry has
been developed in \cite{Mellor-90}. There are neither
purely electromagnetic nor gravitational modes. Indeed, we have four
mixed electromagnetic and gravitational fields, two of them called
polar fields, $Z_{1}^{+}$ and $Z_{2}^{+}$ (since they impart no
rotation to the black hole) and two named axial fields, $Z_{1}^{-}$
and $Z_{2}^{-}$. 
It is possible to express their dynamics in four decoupled wave equations,
two for the axial fields and two for the polar fields. Their deduction
can be found in \cite{Mellor-90} and
references therein. Here we just show the expressions, which will be useful
throughout our work.
                                                                              
The axial perturbations $Z_{1,2}^{-}$ are governed by wave equations which
have the same form as Eq. (\ref{wave_equation}), but with effective potentials
given by
\begin{equation}
V_{1}^{-} = h(r) \left[\frac{\ell (\ell + 1)}{r^2} + \frac{4q^2}{r^2} -
\frac{3m - \sqrt{9m^2 + 8cq^2}}{r^3}\right] \ ,  \label{V-1}
\end{equation}
\begin{equation}
V_{2}^{-} = h(r) \left[\frac{\ell (\ell + 1)}{r^2} + \frac{4q^2}{r^2} -
\frac{3m + \sqrt{9m^2 + 8cq^2}}{r^3}\right] \ ,  \label{V-2}
\end{equation}
respectively, with $2c = (\ell - 1)(\ell + 2)$.
                                                                              
The polar perturbations $Z_{1,2}^{+}$ are subjected to rather
cumbersome potentials, as we can see below:
\begin{equation}
V_{1}^{+}=\frac{h(r)}{r^{3}}\left[ U+\frac{1}{2}(p_{1}-p_{2})W\right] \ ,
\end{equation}
\begin{equation}
V_{2}^{+}=\frac{h(r)}{r^{3}}\left[ U-\frac{1}{2}(p_{1}-p_{2})W\right] \ ,
\end{equation}
\begin{equation}
p_{1}=3m+\sqrt{9m^{2} + 8 c q^{2}} \ ,
\end{equation}
\begin{equation}
p_{2}=3m-\sqrt{9m^{2} + 8 c q^{2}} \ ,
\end{equation}
\begin{equation}
W=\frac{rh(r)}{\bar{\omega}^{2}}(2cr+3m)+\frac{1}{\bar{\omega}}\left( cr+m+%
\frac{2}{3}\Lambda r^{3}\right) \ ,
\end{equation}
\begin{equation}
U=(2cr+3m)W+\left( \bar{\omega}-cr-m-\frac{2}{3}\Lambda r^{3}\right) -\frac{%
2cr^{2}h(r)}{\bar{\omega}}\ ,
\end{equation}
\begin{equation}
\bar{\omega}=cr+3m-\frac{2q^{2}}{r}\ .
\end{equation}                                                  
In the limit $q\rightarrow 0$, the RNdS potentials $V_{2}^{\pm }$ go into the
SdS polar and axial potentials, $V^{\pm
}$. Therefore, the minimum $\ell$ for these fields is $\ell=2$, while the
$V_{1}^{\pm}$ fields admit $\ell=1$ as their minimum $\ell$ value,
since they become electromagnetic perturbations in the limit
$q\rightarrow 0$.

\section{Numerical and semianalytical approaches}

\subsection{Characteristic integration}

In \cite{Gundlach-94} a simple but at the
same time very efficient way of dealing with two-dimensional
d'Alembertians  has been set up. Along the general lines of the
pioneering work
\cite{Price-72}, the authors introduced light-cone
variables $u = t - x$ and $v = t + x$, in terms of which
all the wave equations introduced have the same form. We call $V$ the
generic effective potential and $\psi_{\ell}$ the generic field, and the
equations can be written, in terms of the null coordinates, as
\begin{equation}
-4 \frac{\partial^2}{\partial u \partial v} \psi_{\ell} (u,v) =
V (r(u,v)) \psi_{\ell} (u,v) \ .
\label{uv-eq}
\end{equation}
In the characteristic initial value problem, initial data are specified
on the two null surfaces $u = u_{0}$ and $v = v_{0}$. Since the basic
aspects of the field decay are independent of the initial conditions
(this fact is confirmed by our simulations),
we use
\begin{equation}
\psi_{\ell}(u=u_0,v) = \exp\left[-\frac{(v - v_c)^2}{2\sigma^2}\right] \ ,
\end{equation}
\begin{equation}
\psi_{\ell}(u, v=v_0) = \exp\left[-\frac{(v_0 - v_c)^2}{2\sigma^2}\right] .
\end{equation}
Due to the size of our lattices, the latter constant can be set to zero
for any practical purpose.

Since we do not have analytic solutions to the
time-dependent wave equation with the effective potentials introduced,
one approach is to discretize Eq. (\ref{uv-eq}) and
then implement a finite differencing scheme to solve it
numerically. One possible discretization, used, for example, in
\cite{Wang-01,Brady-97,Brady-99},  is  
\begin{eqnarray}
\lefteqn{\psi_{\ell}(N) = \psi_{\ell}(W) + \psi_{\ell}(E) -
\psi_{\ell}(S) }  \nonumber \\  
& & \mbox{} - \Delta^2 V(S) \frac{ \psi_{\ell}(W) + \psi_{\ell}(E)}{8} +
O(\Delta^4)   \ , 
\label{d-uv-eq}
\end{eqnarray}
where we have used the definitions for the points $N = (u + \Delta, v
+ \Delta)$, $W = (u + \Delta, v)$, $E = (u, v + \Delta)$, and $S =
(u,v)$. With the use of expression 
(\ref{d-uv-eq}), the basic algorithm will cover the  region of
interest in the $u-v$ plane, using the value of the field at three points 
in order to compute it at a fourth.

After the integration is completed, the values $\psi_{\ell}(u_{max}, v)$
and $\psi_{\ell}(u,v_{max})$ are extracted,  where  $u_{max}$ ($v_{max}$)
is  the  maximum value  of $u$ ($v$)  on  the numerical grid. Taking
sufficiently large $u_{max}$ and $v_{max}$, we have good
approximations for the wave function at the event and cosmological
horizons.

\subsection{Noncharacteristic integration}

It is not difficult to set up a numeric algorithm to solve
Eq. (\ref{wave_equation}) with Cauchy data specified
on a $t$ constant surface. We used a fourth order in $x$ and
a second order in $t$ scheme (see, for instance,
 \cite{Levander} for an application of this algorithm to seismic
analysis). The second spatial
derivative at a point $(t,x)$, up to fourth order, is given by
\begin{eqnarray}
\label{spatial}
\psi_{\ell}''(t,x) & = & \frac{1}{12\Delta x^2}
\left[\psi_{\ell}(t,x+2 \Delta x)  -16\psi_{\ell}(t,x+\Delta x)
\right. \nonumber \\
& & \mbox{} + 30\psi_{\ell}(t,x)- 16\psi_{\ell}(t,x-\Delta x)
\nonumber \\
& &  \left. +\psi_{\ell}(t,x-2\Delta x) \right] \ , 
\end{eqnarray}
while the second time derivative up to second order is 
\begin{equation}
\label{time}
\ddot{\psi}_{\ell}(t,x) = \frac{\psi_{\ell}(t+\Delta t,x) -2\psi_{\ell}(t,x) +
\psi_{\ell}(t-\Delta t,x)}{\Delta t^2} \ .
\end{equation}
Given $\psi_{\ell}(t=t_0,x)$ and $\psi_{\ell}=(t=t_0-\Delta t,x)$ 
\linebreak (or $\dot{\psi}_{\ell}(t=t_0,x)$), we can use the discretization in Eqs.
(\ref{spatial}) and (\ref{time}) to solve Eq. (\ref{wave_equation}) and
calculate $\psi_{\ell}(t=t_0+\Delta t,x)$. This is the basic
algorithm. At each interaction, one can control the error by using the
invariant integral (the wave energy) associated with Eq. (\ref{wave_equation})
\begin{equation}
E = \frac{1}{2}\int \left[ \left( \psi_{\ell}'(t,x)\right)^2
+ \left( \dot{\psi}_{\ell}(t,x)\right)^2 +V(x)
\psi_{\ell}(t,x)^2\right] \, dx \ . 
\end{equation}

We make an exhaustive analysis of the asymptotic behavior of the solutions
of Eq. (\ref{wave_equation}) with  initial conditions of the
form
\begin{equation}
\psi_{\ell}(0,x) = \exp\left[ -\frac{(x-x_0)^2}{2\sigma_0^2}\right] \ ,
\end{equation}
\begin{equation}
\label{psidot}
\dot{\psi}_{\ell}(0,x) = A\exp\left[ -\frac{(x-x_1)^2}{2\sigma_1^2}\right]
\ . 
\end{equation}
The results do not depend on the details of the
initial conditions. They are compatible with the ones obtained by the usual
characteristic  integration, with the only, and significant, exception of
the  $\ell=0$ scalar mode. As we will see, its asymptotic value
depends strongly on the initial velocities $\dot{\psi}_{\ell}(0,x)$,
a behavior already advanced in the work \cite{Brady-99}.  

\subsection{WKB analysis}

Considering the Laplace transform of Eq. (\ref{wave_equation}), one
gets the ordinary differential equation   
\begin{equation}
\frac{d^2 \psi_{\ell}(x)}{d x^2}-\left[s^2 + V(x) \right]\psi_{\ell}(x)=0 \ . 
\end{equation}
One finds that there is a discrete set of possible values for $s$
such that the function $\hat{\psi}_{\ell}$, the Laplace-transformed
field, satisfies both boundary conditions,
\begin{equation}
\lim_{x \rightarrow -\infty}\hat{\psi}_{\ell} \, e^{sx}=1 \ ,
\end{equation}
\begin{equation}
\lim_{x \rightarrow +\infty}\hat{\psi}_{\ell} \, e^{-sx}=1 \ .
\end{equation}
By making the formal replacement $s=i\omega$, we have the usual quasinormal
mode boundary conditions. The frequencies $\omega$ (or $s$) are
called quasinormal frequencies.

The semianalytic approach used in this work \cite{Schutz-85,Iyer-87}
is a very efficient algorithm to calculate the quasinormal
frequencies, which have been applied in a variety of situations
\cite{WKB-ap}. With this method, the quasinormal modes are given by
\begin{equation}
\omega_{n}^{2} = \left(V_{0} + P\right) - i\left(n + \frac{1}{2}\right)
            \left(-2V_{0}^{(2)}\right)^{1/2} \left(1 + Q\right)
\label{w-WKB}
\end{equation}
where the quantities $P$ and $Q$ are determined using
\begin{equation}
P = \frac{1}{8} \left[\frac{V_{0}^{(4)}}{V_{0}^{(2)}}\right]
    \left(\frac{1}{4}+\alpha^2\right)
    - \frac{1}{288} \left[\frac{V_{0}^{(3)}}{V_{0}^{(2)}}\right]^{2}
    \left(7 + 60 \alpha^2\right) \ ,
\label{P-WKB}
\end{equation}
\begin{widetext}
\begin{eqnarray}
Q &=& \frac{1}{-2V_{0}^{(2)}} \left\{ 
    \frac{5}{6912} \left[\frac{V_{0}^{(3)}}{V_{0}^{(2)}}\right]^{4}
    \left(77 + 188\alpha^2\right)  
    - \frac{1}{384} \left[\frac{V_{0}^{(3) 2}
    V_{0}^{(4)}}{V_{0}^{(2)}}\right] \left(51 + 100\alpha^2\right)
    \right.    \nonumber  \\ 
   && + \frac{1}{2304} \left[\frac{V_{0}^{(4)}}{V_{0}^{(2)}}\right]^{2}
    \left(67 + 68\alpha^2\right) 
 + \frac{1}{288} \left[\frac{V_{0}^{(3)}
    V_{0}^{(5)}}{V_{0}^{(2) 2}}\right] \left(19 + 28\alpha^2\right) 
   \nonumber  \\
 &&\left. \mbox{} - \frac{1}{288} 
 \left[\frac{V_{0}^{(6)}}{V_{0}^{(2)}}\right] \left(5 + 4\alpha^2\right) 
 \right\} 
\label{Q-WKB}
\end{eqnarray}
\end{widetext}

In Eqs. (\ref{w-WKB})-(\ref{Q-WKB}), $\alpha = n + 1/2$ and
the superscript $(i)$ denotes differentiation, with respect to $x$, of
the potential $V(x)$. The potential and its derivatives are then
calculated at the point $x_{0}$, where $V(x)$ is an extremum.
The integer $n$ labels the modes
\begin{equation}
n = \begin{cases}

    0,1,2,\ldots,     & \rm{Re}(\omega_n)>0  \ ,\\
    -1,-2,-3,\ldots,  & \rm{Re}(\omega_n)<0 \ .
    \end{cases} 
\end{equation}

\section{Near Extreme Limit}

\subsection{Schwarzschild--de Sitter black hole}

To characterize the near extreme limit of the Schwarzschild--de Sitter
geometry, it is  convenient to define the  dimensionless parameter
$\bar{\delta}$: 
\begin{equation}
\bar{\delta}=\sqrt{1 - 9m^2\Lambda} \ .
\end{equation}
The limit $0<\bar{\delta}\ll1$ is the near extreme limit, where the
horizons  are distinct, but very close. In this regime, analytical
expressions for the frequencies have been calculated
\cite{Cardoso-03,Molina-03}. For the scalar and electromagnetic
fields, the quasinormal frequencies are  
\begin{eqnarray}
\omega_n  & = & \left[
\frac{\Lambda}{3} - 3m^{2}\Lambda^{2} \right]^{1/2} \nonumber \\
& & \times \left\{ \left[ 
\ell (\ell + 1) - \frac{1}{4} \right]^{1/2}  - i
\left(n+\frac{1}{2}\right)  \right\} . 
\label{w-qext-sc}
\end{eqnarray}
For the axial and polar gravitational fields, the frequencies are
given by
\begin{eqnarray}
\omega_n  & = & \left[
\frac{\Lambda}{3} - 3m^{2}\Lambda^{2} \right]^{1/2} \nonumber \\
& & \times \left\{ \left[ 
(\ell + 2)(\ell - 1) - \frac{1}{4} \right]^{1/2}  - i
\left(n+\frac{1}{2}\right)  \right\} .  \nonumber\\
\label{w-qext-gr}
\end{eqnarray}
They can be compared with the numerical and semianalytic methods
presented in the previous section.      

\begin{figure}
\resizebox{1\linewidth}{!}{\includegraphics*{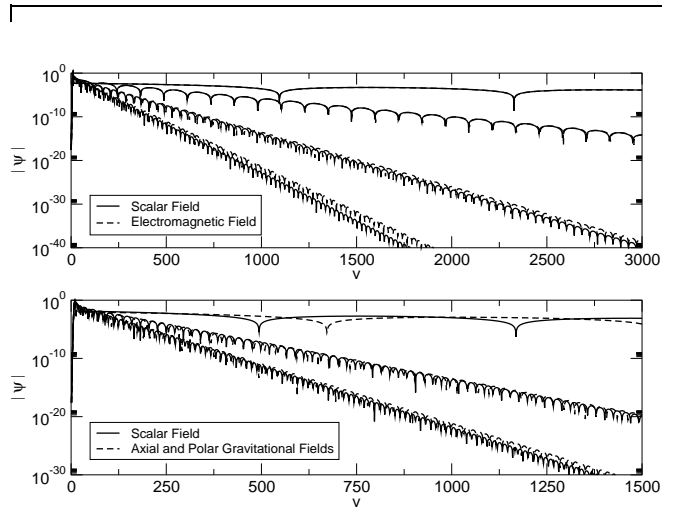}}
\caption{Decay of the scalar and electromagnetic fields with
$\ell=1$, and of the scalar, axial, and polar gravitational fields,
with $\ell=2$, with the SdS geometry approaching the near extreme
limit.The parameters for the geometry are $m=1.0$ and
$\bar{\delta}=0.01,0.1,0.3$. }    
\label{near-ext-l1-2}
\end{figure}

Direct calculation of the wave functions confirms that, in the near
extreme limit, their dynamics is simple, with the late-time decay of the
fields being dominated by quasinormal modes. 
All the types of perturbation tend to coincide near the extreme limit.
In addition, as we approach the extreme limit, the oscillation period increases
and the exponential decay rate decreases.
These conclusions, illustrated in Fig. \ref{near-ext-l1-2} for
$\ell = 1,2$, are  consistent with the ones presented in
\cite{Cardoso-03,Molina-03}. 

\begin{figure}
\resizebox{1\linewidth}{!}{\includegraphics*{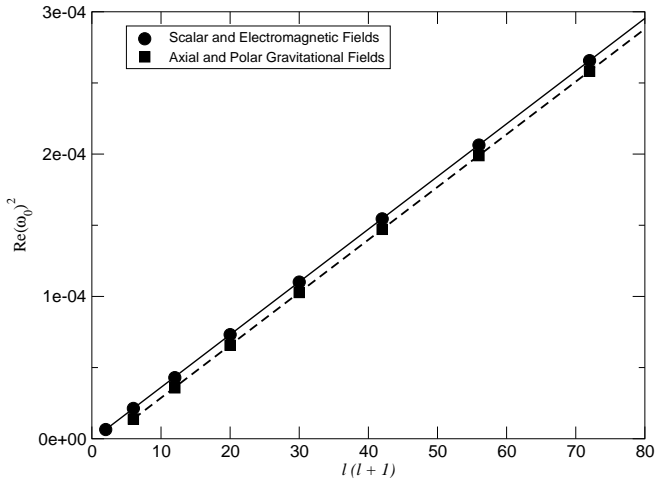}}
\caption{Curve of $\textrm{Re}(\omega_0)^{2}$ with
$\ell(\ell + 1)$, for the scalar, electromagnetic and gravitational
fields, in the near extreme limit. The dots are the numerical
results and the solid lines represent the analytical results.  The
parameters for the geometry are $m=1.0$ and $\bar{\delta}=0.01$.}  
\label{near-ext}
\end{figure}

By using a nonlinear fitting based on a $\chi^2$ 
analysis, it is possible to estimate the real and
imaginary parts of the $n=0$ quasinormal mode. 
These results can be compared with the analytical expressions in
the near extreme cases. In Fig. \ref{near-ext}, we analyze
the dependence of the frequencies on $\ell$. The accordance between
analytic and numerical data is extremely good.

\subsection{Reissner--Nordstr\"{o}m--de Sitter black hole}

For the RNdS case, the near extreme limit corresponds to the
situation where the event and cosmological horizons are very
close to each other. It is natural to define the dimensionless parameter $%
\delta$ as
\begin{equation}
\delta=\frac{r_{c}-r_{+}}{r_{+}} \ ,
\end{equation}
where $0 < \delta\ll 1$. In this limit the dynamics can be
analytically characterized, as has been analyzed in
\cite{Cardoso-03}. More general settings, including RNdS geometries,
were explored in \cite{Molina-03}.
                                                                              
The function $h(x) \equiv h(r(x))$ can be analytically calculated
\cite{Cardoso-03,Molina-03}, with the result
\begin{equation}
h(x) = \frac{(r_{c} - r_{+})\kappa_{+}}{2\cosh^{2}(\kappa_{+}x)} +
O\left(\delta^{3}\right) \ .  \label{hx}
\end{equation}
We have five different fields at hand: the scalar field ($Z_{sc}$), two axial
fields ($Z_{1}^{-}$, $Z_{2}^{-}$), and two polar fields ($Z_{1}^{+}$, $%
Z_{2}^{+}$). For each one, we have a different potential. In the
near extreme limit, we have
\begin{equation}
V(x) = \Omega(r_{+}) h(x) + O\left( \delta \right) = \frac{V_{0}}{
\cosh^{2}\left(\kappa_{+}x\right)} + O\left( \delta \right) \ .
\end{equation}
The constant $V_{0}$ in the scalar case is denoted by $V_{0}^{sc}$, in the
axial cases by $\{ V_{0}^{1-}, V_{0}^{2-}\}$, and in the polar cases by $\{
V_{0}^{1+}, V_{0}^{2+}\}$. The foregoing expression is a P\"{o}schl-Teller
potential \cite{Poschl-33}.
                                                                              
For the scalar field, $V_{0}^{sc}$ has been calculated in
\cite{Molina-03}, and is given by
\begin{equation}
V_{0}^{sc} = \frac{\ell(\ell+1) (r_{c} - r_{+}) \kappa_{+} }{2r_{+}^2} \ .
\end{equation}
We proceed to the analysis of the coupled electromagnetic and
gravitational fields. We take the analytical
expressions for all potentials $V_{1,2}^{\pm}$ and
we go to the near extreme limit. For the two
axial potentials, we have
\begin{equation}
V_{0}^{1-} = \frac{(r_{c} - r_{+})\kappa_{+}}{2 r_{+}^4} \left[ \ell (\ell +
1)r_{+}^2 + 4q^2 - r_{+} S_{1} \right] \ ,
\end{equation}
\begin{equation}
V_{0}^{2-} = \frac{(r_{c} - r_{+})\kappa_{+}}{2 r_{+}^4} \left[ \ell (\ell +
1)r_{+}^2 + 4q^2 - r_{+} S_{2} \right] \ ,
\end{equation}
where
\begin{equation}
S_{1} = 3m - \sqrt{ 9m^2 + 4(\ell + 2)(\ell - 1)q^2} \ ,
\end{equation}
\begin{equation}
S_{2} = 3m + \sqrt{ 9m^2 + 4(\ell + 2)(\ell - 1)q^2} \ .
\end{equation}
                                                                              
We can now turn to the two polar potentials. The constants
$V_{0}^{1+}$ and $V_{0}^{2+}$ are given by
\begin{widetext}
\begin{equation}
V_{0}^{1+} = \frac{\kappa_{+}(r_{c}-r_{+})}{2r_{+}^4} \left[ \frac{\left(
2cr_{+}^2 + 3mr_{+} + r_{+}\sqrt{9m^2+8cq^2} \right) \left(cr_{+} + m
+ \frac{2\Lambda r_{+}^3}{3} \right)}{cr_{+} + 3m -\frac{2q^2}{r_{+}}}
+ C \right] \  ,
\end{equation}
\begin{equation}
V_{0}^{2+} = \frac{\kappa_{+}(r_{c}-r_{+})}{2r_{+}^4} \left[ \frac{\left(
2cr_{+}^2 + 3mr_{+} - r_{+}\sqrt{9m^2+8cq^2} \right) \left(cr_{+} + m
+ \frac{2\Lambda r_{+}^3}{3} \right)}{cr_{+} + 3m -\frac{2q^2}{r_{+}}}
+ C \right] \ ,
\end{equation}
\end{widetext}
with $C = 2mr_{+} - 2q^2 - 2 \Lambda r_{+}^4 /3$ and $2c=(\ell
+ 2)(\ell - 1)$.
The quasinormal modes associated with the P\"{o}schl-Teller potential have
been extensively studied \cite{Ferrari-84,Beyer-99}. The frequencies
$\omega_{n}$ are given by
\begin{equation}
\frac{\omega_{n}}{\kappa_{+}} = \sqrt{\frac{V_{0}}{\kappa_{+}^2} -
\frac{1}{4} } - i \left( n + \frac{1}{2} \right) \ ,  \label{qnfreq}
\end{equation}
with $n \in \{0,1,\ldots\}$ labeling the modes. Using expression (\ref
{qnfreq}) and the expression for $V_{0}$, the frequencies can be
easily calculated.
                                                                              
\begin{figure}
\resizebox{1\linewidth}{!}{\includegraphics*{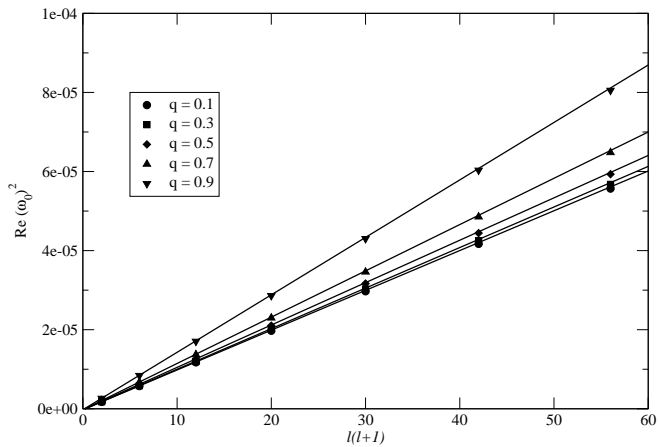}}
\caption{Near extreme fundamental frequencies for different
values of the charge $q$, for the lower multipole
mode of the scalar field. Analytical values are represented by
straight lines, and numerical values appear as dots. The parameters
for the geometry are $m=1.0$ and $\delta=10^{-3}$.}
\label{ES-q-ext}
\end{figure}

\begin{figure}
\resizebox{1\linewidth}{!}{\includegraphics*{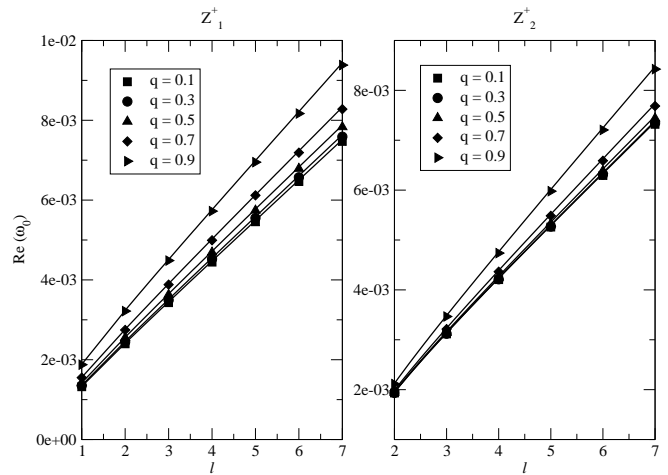}}
\caption{Near extreme fundamental frequencies for different
values of the charge $q$, for the lower multipole
mode of the polar fields. On the left are the
data for $Z_{1}^{+}$ and, on the left, for $Z_{2}^{+}$. The parameters
for the geometry are $m=1.0$ and $\delta=10^{-3}$. }
\label{Z+12-q-ext}
\end{figure}

\begin{figure}
\resizebox{1\linewidth}{!}{\includegraphics*{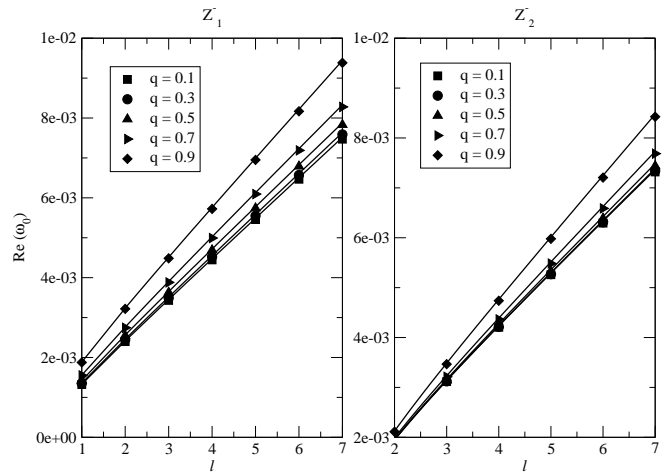}}
\caption{Near extreme fundamental frequencies
for different values of the charge $q$, for the lower multipole
mode of the axial fields. The data for
$Z_{1}^{-}$ appear on the left, and those for $Z_{2}^{-}$, on the
right. The parameters for the geometry are $m=1.0$ and $\delta=10^{-3}$.}
\label{Z-12-q-ext}
\end{figure}

We can also use the numerical method to analyze the field decay in the
near extreme limit. Using a nonlinear fitting based on
$\chi^2$ analysis  for the wave
functions, we can estimate the quasinormal frequencies. These results can be
compared with the analytical expressions in the near extreme cases. The
accordance between both sets of results is extremely good. We illustrate
this point in Figs. \ref{ES-q-ext}--\ref{Z-12-q-ext}.

Direct calculation of the wave functions confirms that, in the near extreme
limit, the dynamics of the fields is simple, with the late-time decay being
completely dominated by quasinormal modes.
                                                                              

\section{Intermediary Region in Parameter Space}

\subsection{Schwarzschild--de Sitter black hole}

\subsubsection{Scalar field with $\ell = 0$}

\begin{figure}
\resizebox{1\linewidth}{!}{\includegraphics*{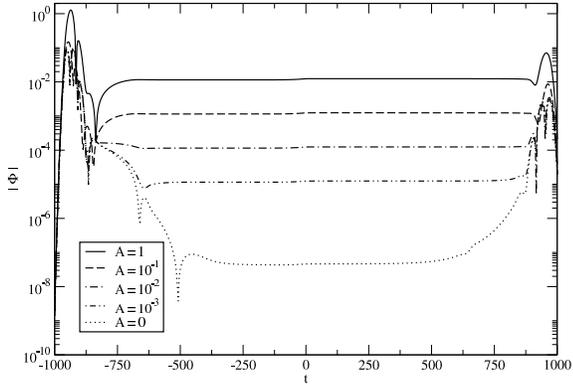}}
\caption{Asymptotic solutions $\phi(x,t)$ obtained
by noncharacteristic numerical integration with $\ell=0$, $\Lambda=10^{-4}$
and $m=1.0$. The curves correspond to different values of $A$ in
the initial condition (\ref{psidot}).}  
\label{non-char}
\end{figure}

Only scalar perturbations can have zero total angular momentum.
Solutions of  Eq. (\ref{wave_equation}) with $\ell = 0$ lead to  a
constant tail, as already shown in \cite{Brady-97,Brady-99}. 
This is confirmed
in Fig. \ref{non-char}. 
The novelty here is the dependence of the asymptotic value
on the $\dot{\psi}_{\ell}(0,x)$ initial condition. 
Figure \ref{non-char} reveals the appearance
of the constant value $\phi_0$ for large $t$, and its dependence on
$\dot{\psi}_{\ell}(0,x)$. Note that $\phi_0$ falls below 
$10^{-7}$ for $\dot{\psi}_{\ell}(0,x)=0$. These results are in
accordance with the analytical predictions of \cite{Brady-99}, which
give
\begin{equation}
\psi \left(\infty,r\right) = \frac{r}{r_c^2}\int_0^{r_c}
\dot{\psi}(0,s)s\frac{ds}{h'(s)}.
\end{equation}

\subsubsection{Fields with $\ell > 0$ }

We can have scalar and vector fields with angular momentum
$\ell=1$, and with $\ell > 1$, it is possible to introduce 
gravitational fields also.  Their behavior is described
in general by three phases. The first
corresponds to the quasinormal modes generated from the presence of the
black hole itself. A little later there is a region of power law decay,
which continues indefinitely in an asymptotic flat space. In the presence of a
positive cosmological constant, however, an exponential decay takes
over in the latest period.

\begin{figure}
\resizebox{1\linewidth}{!}{\includegraphics*{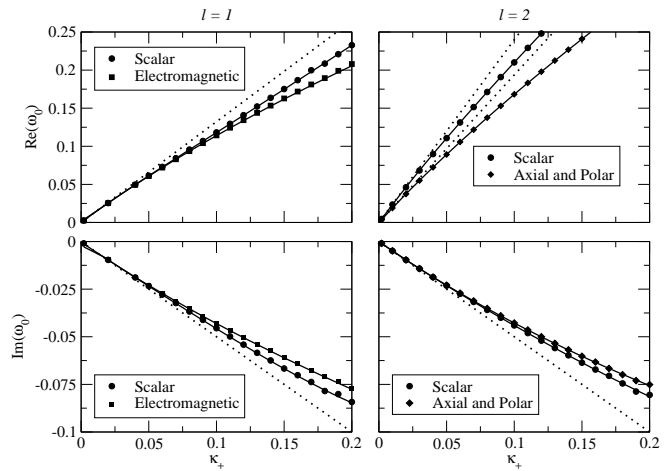}}
\caption{Graphs of the real and imaginary part of the
fundamental frequencies $(n=0)$ with $\kappa_{+}$. The dotted
lines are the near extreme results, the dots are the numerical
results and the continuous curves are the semianalytic results. In
the graphs, $m=1.0$.}
\label{w-kappap}
\end{figure}

As the separation of the horizons increases, the quasinormal
frequencies deviate from those predicted by expressions
(\ref{w-qext-sc}) and (\ref{w-qext-gr}). In Fig. \ref{w-kappap},
this is illustrated for $\ell=1,2$. Some qualitatively different
effects show up when we turn away from the near extreme limit. For a
small cosmological constant the asymptotic behavior is dominated by an
exponentially decaying mode rather than by a quasinormal mode, for all
perturbations considered. 

It is interesting to compare the values obtained for the fundamental
modes using the numerical and semianalytic methods. We find that the
agreement between them is good, for the whole range of $\Lambda$. The
difference is smaller for the first values of $\ell$. 
This is expected, since the numerical calculations work better in this
region.  In Table \ref{SC-num-WKB}, we illustrate
these observations with a few values of $\Lambda$. It is important to
mention that quasinormal frequencies for the SdS black hole were
already calculated in a recent paper \cite{Zhidenko}, by applying a
variation of the WKB method used here \cite{Konoplya-03}. There are earlier
papers calculating quasinormal modes in this geometry, for example,
\cite{dsqnm}.   

\begin{table*}
\caption{Fundamental frequencies for the scalar
field in SdS, obtained using the numerical and semianalytical
methods. In this table, $m=1.0$.}
\begin{ruledtabular}
\begin{tabular}{cccccc}

\multicolumn{2}{c}{q = 0}         &
\multicolumn{2}{c}{Numerical}&
\multicolumn{2}{c}{Semianalytical}\\ 

$\ell$ & $\Lambda$ &Re($\omega_0$)& -Im($\omega_0$)&Re($\omega_0$)& -Im($\omega_0$)\\
\hline   
1  & $10^{-5}$ & 2.930$\times 10^{-1}$ & 9.753$\times 10^{-1}$ & 2.911$\times 10^{-1}$ & 9.780$\times 10^{-2}$ \\
 
   & $10^{-4}$ & 2.928$\times 10^{-1}$ & 9.764$\times 10^{-2}$ & 2.910$\times 10^{-1}$ & 9.797$\times 10^{-2}$ \\
 
   & $10^{-3}$ & 2.914$\times 10^{-1}$ & 9.726$\times 10^{-2}$ & 2.896$\times 10^{-1}$ & 9.771$\times 10^{-2}$ \\
 
   & $10^{-2}$ & 2.770$\times 10^{-1}$ & 9.455$\times 10^{-2}$ & 2.753$\times 10^{-1}$ & 9.490$\times 10^{-2}$ \\
 
   & $10^{-1}$ & 8.159$\times 10^{-2}$ & 3.123$\times 10^{-2}$ & 8.144$\times 10^{-2}$ & 3.137$\times 10^{-2}$ \\
\hline 
2  &$10^{-5}$ & 4.840$\times 10^{-1}$ & 9.653$\times 10^{-2}$ & 4.832$\times 10^{-1}$ & 9.680$\times 10^{-2}$ \\
 
   & $10^{-4}$ & 4.833$\times 10^{-1}$ & 8.948$\times 10^{-2}$ & 4.830$\times 10^{-1}$ & 9.677$\times 10^{-2}$ \\
 
   & $10^{-3}$ & 4.816$\times 10^{-1}$ & 8.998$\times 10^{-2}$ & 4.809$\times 10^{-1}$ & 9.643$\times 10^{-2}$ \\ 
 
   & $10^{-2}$ & 4.598$\times 10^{-1}$ & 8.880$\times 10^{-2}$ & 4.592$\times 10^{-1}$ & 9.290$\times 10^{-2}$ \\
 
   & $10^{-1}$ & 1.466$\times 10^{-1}$ & 3.068$\times 10^{-2}$ & 1.466$\times 10^{-1}$ & 3.070$\times 10^{-2}$ \\
\hline  
3  &$10^{-5}$ & 6.769$\times 10^{-1}$ & 8.662$\times 10^{-2}$ & 6.752$\times 10^{-1}$ & 9.651$\times 10^{-2}$ \\
 
   & $10^{-4}$ & 6.754$\times 10^{-1}$ & 8.654$\times 10^{-2}$ & 6.749$\times 10^{-1}$ & 9.647$\times 10^{-2}$ \\
 
   & $10^{-3}$ & 6.732$\times 10^{-1}$ & 8.660$\times 10^{-2}$ & 6.720$\times 10^{-1}$ & 9.611$\times 10^{-2}$ \\
 
   & $10^{-2}$ & 6.437$\times 10^{-2}$ & 9.200$\times 10^{-2}$ & 6.428$\times 10^{-2}$ & 9.235$\times 10^{-2}$ \\
 
   & $10^{-1}$ & 2.091$\times 10^{-2}$ & 3.054$\times 10^{-2}$ & 2.091$\times 10^{-2}$ & 3.056$\times 10^{-2}$ \\

\end{tabular}
\end{ruledtabular}
\label{SC-num-WKB}
\end{table*}

\begin{table*}
\caption{Fundamental frequencies for the
electromagnetic field in SdS, obtained using the numerical and
semianalytical methods.  In this table, $m=1.0$.}  
\begin{ruledtabular}
\begin{tabular}{cccccc}

\multicolumn{2}{c}{q = 0}         &
\multicolumn{2}{c}{Numerical}&
\multicolumn{2}{c}{Semianalytical}\\

$\ell$ & $\Lambda$ &Re($\omega_0$)& -Im($\omega_0$)&Re($\omega_0$)& -Im($\omega_0$)\\
\hline 
1  &$10^{-5}$ & 2.481$\times 10^{-1}$ & 9.226$\times 10^{-2}$ & 2.459$\times 10^{-1}$ & 9.310$\times 10^{-2}$ \\
 
   & $10^{-4}$ & 2.481$\times 10^{-1}$ & 9.223$\times 10^{-2}$ & 2.457$\times 10^{-1}$ & 9.307$\times 10^{-2}$ \\
 
   & $10^{-3}$ & 2.475$\times 10^{-1}$ & 9.176$\times 10^{-2}$ & 2.448$\times 10^{-1}$ & 9.270$\times 10^{-2}$ \\
 
   & $10^{-2}$ & 2.374$\times 10^{-1}$ & 8.839$\times 10^{-2}$ & 2.352$\times 10^{-1}$ & 8.896$\times 10^{-2}$ \\
 
   & $10^{-1}$ & 8.035$\times 10^{-2}$ & 3.027$\times 10^{-2}$ & 8.023$\times 10^{-2}$ & 3.033$\times 10^{-2}$ \\
\hline 
2  &$10^{-5}$ & 4.577$\times 10^{-1}$ & 8.985$\times 10^{-2}$ & 4.571$\times 10^{-1}$ & 9.506$\times 10^{-2}$ \\
 
   & $10^{-4}$ & 4.575$\times 10^{-1}$ & 8.991$\times 10^{-2}$ & 4.569$\times 10^{-1}$ & 9.502$\times 10^{-2}$ \\
 
   & $10^{-3}$ & 4.559$\times 10^{-1}$ & 9.439$\times 10^{-2}$ & 4.551$\times 10^{-1}$ & 9.464$\times 10^{-2}$ \\ 
 
   & $10^{-2}$ & 4.371$\times 10^{-1}$ & 8.941$\times 10^{-2}$ & 4.364$\times 10^{-1}$ & 9.074$\times 10^{-2}$ \\
 
   & $10^{-1}$ & 1.458$\times 10^{-1}$ & 3.037$\times 10^{-2}$ & 1.458$\times 10^{-1}$ & 3.038$\times 10^{-2}$ \\
\hline  
3  &$10^{-5}$ & 6.578$\times 10^{-1}$ & 8.365$\times 10^{-2}$ & 6.567$\times 10^{-1}$ & 9.563$\times 10^{-2}$ \\
 
   & $10^{-4}$ & 6.576$\times 10^{-1}$ & 8.349$\times 10^{-2}$ & 6.564$\times 10^{-1}$ & 9.559$\times 10^{-2}$ \\
 
   & $10^{-3}$ & 6.547$\times 10^{-1}$ & 8.399$\times 10^{-2}$ & 6.538$\times 10^{-1}$ & 9.520$\times 10^{-2}$ \\
 
   & $10^{-2}$ & 6.276$\times 10^{-2}$ & 8.852$\times 10^{-2}$ & 6.267$\times 10^{-1}$ & 9.125$\times 10^{-2}$ \\
   
   & $10^{-1}$ & 2.085$\times 10^{-2}$ & 3.039$\times 10^{-3}$ & 2.085$\times 10^{-1}$ & 3.040$\times 10^{-2}$ \\

\end{tabular}
\end{ruledtabular}
\label{EL-num-WKB}
\end{table*}

\begin{table*}
\caption{Fundamental frequencies for the axial and polar
gravitational fields in SdS, obtained using the numerical and semianalytical
methods.  In this table, $m=1.0$.} 
\begin{ruledtabular}
\begin{tabular}{cccccc}

\multicolumn{2}{c}{q = 0}         &
\multicolumn{2}{c}{Numerical}&
\multicolumn{2}{c}{Semianalytical}\\

$\ell$ & $\Lambda$ &Re($\omega_0$)& -Im($\omega_0$)&Re($\omega_0$)&
-Im($\omega_0$)\\ 
\hline 
2  &$10^{-5}$ & 3.738$\times 10^{-1}$ & 8.883$\times 10^{-2}$ & 3.731$\times 10^{-1}$ & 8.921$\times 10^{-2}$ \\

   & $10^{-4}$ & 3.737$\times 10^{-1}$ & 8.880$\times 10^{-2}$ & 3.730$\times 10^{-1}$ & 8.918$\times 10^{-2}$ \\
 
   & $10^{-3}$ & 3.721$\times 10^{-1}$ & 8.850$\times 10^{-2}$ & 3.715$\times 10^{-1}$ & 8.888$\times 10^{-2}$ \\ 
 
   & $10^{-2}$ & 3.566$\times 10^{-1}$ & 8.538$\times 10^{-2}$ & 3.560$\times 10^{-1}$ & 8.572$\times 10^{-2}$ \\
 
   & $10^{-1}$ & 1.179$\times 10^{-1}$ & 3.020$\times 10^{-2}$ & 1.179$\times 10^{-1}$ & 3.023$\times 10^{-2}$ \\
\hline 
3  &$10^{-5}$ & 5.999$\times 10^{-1}$ & 8.677$\times 10^{-2}$ & 5.992$\times 10^{-1}$ & 9.272$\times 10^{-2}$ \\
 
   & $10^{-4}$ & 5.996$\times 10^{-1}$ & 8.676$\times 10^{-2}$ & 5.990$\times 10^{-1}$ & 9.269$\times 10^{-2}$ \\
 
   & $10^{-3}$ & 5.972$\times 10^{-1}$ & 8.971$\times 10^{-2}$ & 5.966$\times 10^{-1}$ & 9.234$\times 10^{-2}$ \\
 
   & $10^{-2}$ & 5.725$\times 10^{-1}$ & 8.695$\times 10^{-2}$ & 5.718$\times 10^{-1}$ & 8.874$\times 10^{-2}$ \\
 
   & $10^{-1}$ & 1.900$\times 10^{-1}$ & 3.030$\times 10^{-2}$ & 1.900$\times 10^{-2}$ & 3.032$\times 10^{-2}$ \\
\hline  
4  &$10^{-5}$ & 8.106$\times 10^{-1}$ & 8.810$\times 10^{-2}$ & 8.091$\times 10^{-1}$ & 9.417$\times 10^{-2}$ \\
 
   & $10^{-4}$ & 8.102$\times 10^{-1}$ & 8.781$\times 10^{-2}$ & 8.087$\times 10^{-1}$ & 9.413$\times 10^{-2}$ \\
 
   & $10^{-3}$ & 8.070$\times 10^{-1}$ & 8.799$\times 10^{-2}$ & 8.055$\times 10^{-1}$ & 9.376$\times 10^{-2}$ \\
 
   & $10^{-2}$ & 7.733$\times 10^{-2}$ & 8.714$\times 10^{-2}$ & 7.720$\times 10^{-1}$ & 9.000$\times 10^{-2}$ \\
 
   & $10^{-1}$ & 2.564$\times 10^{-2}$ & 3.034$\times 10^{-3}$ & 2.563$\times 10^{-1}$ & 3.036$\times 10^{-2}$ \\

\end{tabular}
\end{ruledtabular}
\label{AX-num-WKB}
\end{table*}

\begin{figure}
\resizebox{1\linewidth}{!}{\includegraphics*{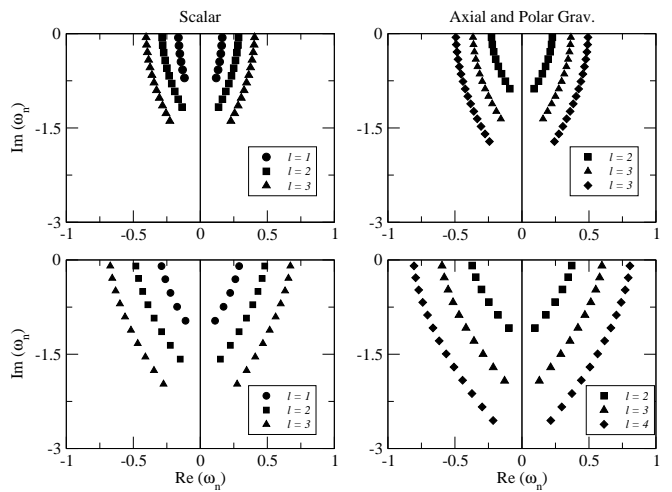}}
\caption{Quasinormal modes of the scalar, axial and polar
gravitational fields, for higher modes. The parameter for the curves
are $m=1.0$ and $\Lambda=10^{-3}$.}
\label{multi-n}
\end{figure}

The first higher $n$ modes cannot be obtained from the numerical
solution, but can be calculated by the semianalytical method. As the
cosmological constant decreases, the real and imaginary parts of the
frequencies increase, up to the limit where  the geometry is
asymptotically flat. The behavior of the modes is illustrated in
Fig. \ref{multi-n}. The behavior of the electromagnetic field is
similar.   

\begin{figure}
\setlength{\unitlength}{1.0mm}
\resizebox{1\linewidth}{!}{\includegraphics*{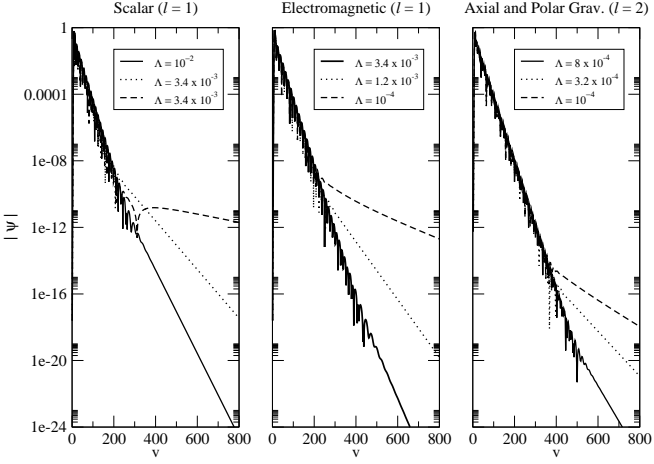}}
\caption{Exponential tail for the scalar and
electromagnetic fields with $\ell=1$, and for the axial and polar
gravitational field with $\ell=2$. In the graphs, $m=1.0$. }
\label{exp-dec}
\end{figure}

A $\chi^2$ analysis of the data presented in 
Figure \ref{exp-dec} shows that the massless scalar, electromagnetic,
and gravitational perturbations in SdS geometry behave as
\begin{eqnarray}
\psi_{\ell}^{sc}\sim e^{-k_{exp}^{sc}t} & {\rm with} &
t\rightarrow\infty  \ ,
\end{eqnarray}
\begin{eqnarray}
\psi_{\ell}^{el}\sim e^{-k_{exp}^{el}t} & {\rm with} &
t\rightarrow\infty  \ ,
\end{eqnarray}
\begin{eqnarray}
\psi_{\ell}^{ax}\sim e^{-k_{exp}^{ax}t} & {\rm with} &
t\rightarrow\infty  \ ,
\end{eqnarray}
\begin{eqnarray}
\psi_{\ell}^{po}\sim e^{-k_{exp}^{po}t} & {\rm with} &
t\rightarrow\infty
\end{eqnarray}
for $t$ sufficiently large. At the event and the cosmological horizons,
$t$ is substituted, respectively, by  $v$ and $u$.

\begin{figure}
\setlength{\unitlength}{1.0mm}
\resizebox{1\linewidth}{!}{\includegraphics*{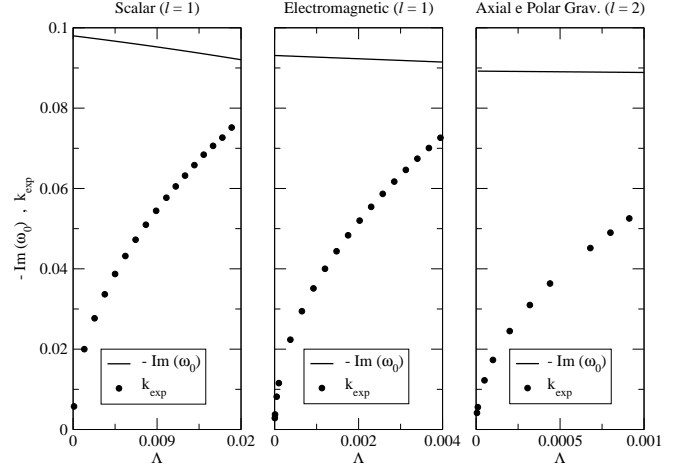}}
\caption{Approaching of  $-\textrm{Im}
(\omega_0)$ and $k_{exp}$, for the scalar, electromagnetic and
gravitational fields, in the SdS geometry. Above a certain critical
value of $\Lambda$ (roughly $1.7 \times 10^{-2}$, $4.0 \times
10^{-3}$ and $1.2 \times 10^{-3}$ respectively, for the parameters used in
the graphs), a tail it is not observed. For all curves, the mass is set to
$m=1.0$. }      
\label{kexp-wI}
\end{figure}

The numerical simulations developed in the present work reveal an
interesting transition between oscillatory modes and exponentially decaying
modes. As shown in Fig. \ref{kexp-wI}, as the cosmological
constant increases, the absolute value of  $-\textrm{Im}
(\omega_0)$ decreases. 

Above a certain critical value of $\Lambda$ we do not observe the
exponential tail, since the coefficient  $k_{exp}$ is larger than
$-\textrm{Im}(\omega_0)$ and thus the decaying quasinormal mode
dominates. But for $\Lambda$ smaller than this critical value,
$-\textrm{Im}(\omega_0)$ turns out to be larger than $k_{exp}$, and the
exponential tail dominates. Certainly, for a small enough cosmological
constant, the exponential tail dominates in the various cases
considered here.

\begin{figure}   
\resizebox{1\linewidth}{!}{\includegraphics*{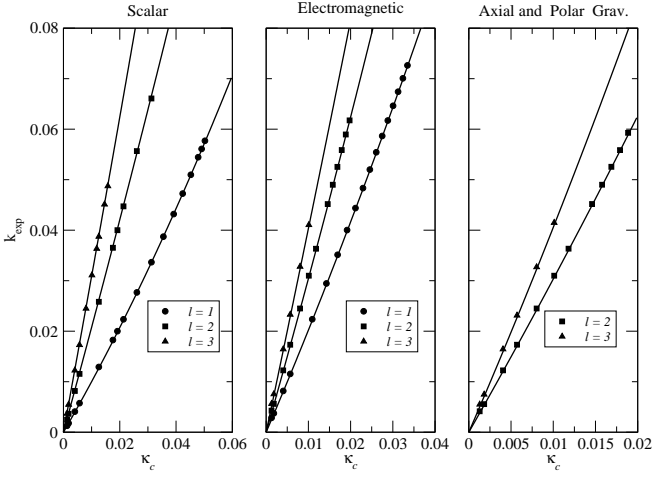}}
\caption{Dependence of  $k_{exp}$ with
$\kappa_{c}$ and $\ell$, in the SdS geometry.  The symbols indicate
the numerical values, and the solid lines are the appropriate fittings. 
For the left graph: 
$k_{exp}^{sc}=1.077\times 10^{-4} + 0.984\kappa_{c} + 3.545\kappa_{c}^2$,  
$k_{exp}^{sc}=-1.863\times 10^{-4} + 2.010\kappa_{c} + 2.608\kappa_{c}^2$ and 
$k_{exp}^{sc}=-1.959\times 10^{-4} + 3.028\kappa_{c} + 2.978\kappa_{c}^2$. 
For the center graph: 
$k_{exp}^{el} = 2.082 \times 10^{-4} + 1.988\kappa_{c} + 6.141\kappa_{c}^2$,  
$k_{exp}^{el} = 2.712 \times 10^{-4} + 2.974\kappa_{c} + 8.284\kappa_{c}^2$ and 
$k_{exp}^{el} = 3.737 \times 10^{-4} + 3.984\kappa_{c} + 4.517\kappa_{c}^2$. 
For the right graph: 
$k_{exp}^{ax} = 2.616 \times 10^{-4}+ 2.974\kappa_{c} +
9.895\kappa_{c}^2$ and
$k_{exp}^{ax} = 4.484\times 10^{-4} + 3.896\kappa_{c} +
18.92\kappa_{c}^2$. In the graphs, $m=1.0$.}
\label{kexp-kappac}
\end{figure}

Another aspect worth mentioning in the intermediate region is the
dependence of the parameters  $k_{exp}^{sc}$, $k_{exp}^{el}$,
$k_{exp}^{ax}$, and $k_{exp}^{po}$ on  $\ell$ and $\kappa_{c}$. The
results suggest that the $k_{exp}$ are at least second differentiable
functions of $\kappa_{c}$. Therefore, close to $\kappa_{c}=0$, we
approximate 
\begin{equation}
k_{exp}^{sc}(\kappa_{c})\approx\ell \left(\kappa_{c}+c^{sc}
\kappa_{c}^{2}\right) \ ,
\end{equation}
\begin{equation}
k_{exp}^{el}(\kappa_{c}) \approx k_{exp}^{ax}(\kappa_{c}) \approx 
k_{exp}^{po}(\kappa_{c}) \approx(\ell+1)
\left(\kappa_{c}+c^{e-g}\kappa_{c}^{2}\right) \ . 
\end{equation}
Previous results are illustrated in Fig. \ref{kexp-kappac}. 

\subsection{Reissner--Nordstr\"{o}m--de Sitter black hole}

We assess here the behavior of the fields in RNdS exterior
geometries that are not near extreme, nor close to the asymptotically
flat limit. Direct numerical simulations and semianalytical (WKB)
methods were largely employed to characterize the fields in this
region.
 
For scalar perturbations with $\ell = 0$, the effective potential is
not positive definite. As already shown in \cite{Brady-97,Brady-99},
solutions of  Eq. (\ref{wave_equation}) with $\ell = 0$ lead to  a
constant tail. It was observed that in the SdS geometry
there is a dependence of the asymptotic value on the
$\dot{\psi}_{\ell}(0,x)$ initial condition, in the context of a Cauchy
type initial value problem. We have checked that the introduction of
electric charge does not alter this picture.

\begin{figure}
\resizebox{1\linewidth}{!}{\includegraphics*{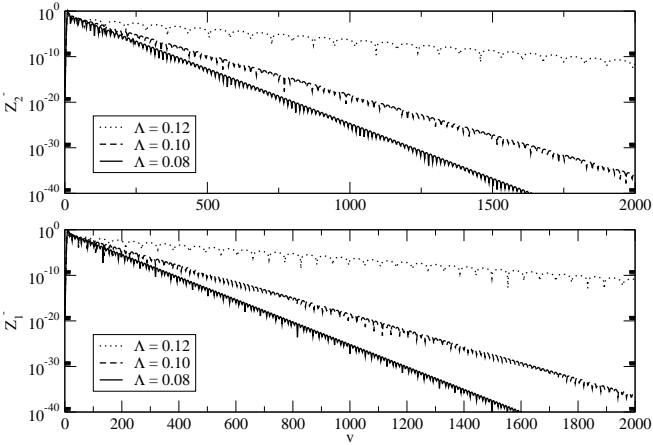}}
\caption{Quasinormal modes for RNdS
$Z_{1}^{-}$ and $Z_{2}^{-}$ fields. The parameters for the geometry are
$q=0.5$ and $m=1.0$. We have used $\ell=2$. The results are similar
for the other fields considered.}
\label{quasinormal-decay}
\end{figure}

If $\ell > 0$, we introduce the $Z_{1}^{\pm}$ fields, and for $\ell
> 1$ the $Z_{2}^{\pm}$ fields can also be analyzed. The first point studied
is the quasinormal phase. If the cosmological constant is high enough,
the decay is  dominated by quasinormal modes, even when they
no longer are accurately predicted by the expressions (\ref{qnfreq}). This
scenario, illustrated in Fig. \ref{quasinormal-decay}, is valid for
all fields considered, with any  charge smaller than its critical
value.
                                                                              
\begin{figure}
\resizebox{1\linewidth}{!}{\includegraphics*{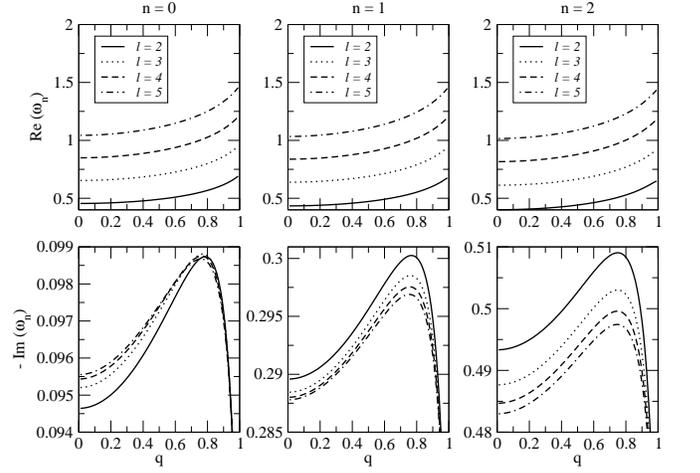}}
\caption{Dependence of $Z_{1}^{+}$ field
frequencies on $q$, for both the real  and imaginary parts, in RNdS. The
results are qualitatively similar for the other fields considered. The
parameters are $m=1.0$ and $\Lambda = 10^{-4}$.}
\label{Re_Im}
\end{figure}
                                                                              
We have observed  that the influence of the electric charge is
mild, although not trivial. The range of variation of the quasinormal
modes with the charge is not very large. We illustrate this point in
Fig. \ref{Re_Im}. It is interesting to compare the values
obtained for the fundamental modes using the numerical and
semianalytical methods. We find a very good agreement between these
results. The difference is smaller for the first values of
$\ell$. This is expected, since the numerical calculations work better
in this region. In Tables
\ref{SC-num-WKB-Z1-}-\ref{SC-num-WKB-Z2+}, we illustrate these
observations for a few values of $\Lambda$ and $q$.
                                                                              
\begin{table*}[tbp]
\caption{Fundamental frequencies for the
$Z_{1}^{-}$ field in RNdS, with
$q=0.5$ and $m=1.0$, using the numerical and semianalytical
      methods.}
\begin{ruledtabular}
\begin{tabular}{cccccc}

\multicolumn{2}{c}{q = 0.5} & \multicolumn{2}{c}{Numerical} &
\multicolumn{2}{c}{Semianalytical} \\ 
$\ell$ & $\Lambda$ & Re($\omega_0$) & -Im($\omega_0$) & Re($\omega_0$) & -Im($%
\omega_0$) \\ \hline
1 &$10^{-5}$ & 2.71$\times 10^{-1}$ & 9.51$\times 10^{-2}$ & 3.28$\times 10^{-1}$ & 9.53$\times 10^{-2}$ \\ 
& $10^{-4}$ & 2.71$\times 10^{-1}$ & 9.50$\times 10^{-2}$ & 3.28$\times 10^{-1}$ & 9.53$\times 10^{-2}$ \\ 
& $10^{-3}$ & 2.70$\times 10^{-1}$ & 9.47$\times 10^{-2}$ & 3.27$\times 10^{-1}$ & 9.49$\times 10^{-2}$ \\ 
& $10^{-2}$ & 2.60$\times 10^{-1}$ & 9.11$\times 10^{-2}$ & 3.15$\times 10^{-1}$ & 9.14$\times 10^{-2}$ \\ 
& $10^{-1}$ & 1.16$\times 10^{-1}$ & 4.08$\times 10^{-2}$ & 1.40$\times 10^{-1}$ & 4.08$\times 10^{-2}$ \\ \hline
2 &$10^{-5}$ & 4.94$\times 10^{-1}$ & 9.72$\times 10^{-2}$ & 4.94$\times 10^{-1}$ & 9.71$\times 10^{-2}$ \\ 
& $10^{-4}$ & 4.93$\times 10^{-1}$ & 9.72$\times 10^{-2}$ & 4.94$\times 10^{-1}$ & 9.71$\times 10^{-2}$ \\ 
& $10^{-3}$ & 4.91$\times 10^{-1}$ & 9.68$\times 10^{-2}$ & 4.92$\times 10^{-1}$ & 9.67$\times 10^{-2}$ \\ 
& $10^{-2}$ & 4.73$\times 10^{-1}$ & 9.31$\times 10^{-2}$ & 4.73$\times 10^{-1}$ & 9.30$\times 10^{-2}$ \\ 
& $10^{-1}$ & 2.09$\times 10^{-1}$ & 4.10$\times 10^{-2}$ & 2.09$\times 10^{-1}$ & 4.16$\times 10^{-2}$ \\ \hline

3 &$10^{-5}$ & 7.05$\times 10^{-1}$ & 8.95$\times 10^{-2}$ & 7.05$\times 10^{-1}$ & 9.77$\times 10^{-2}$ \\ 
& $10^{-4}$ & 7.05$\times 10^{-1}$ & 8.94$\times 10^{-2}$ & 7.05$\times 10^{-1}$ & 9.76$\times 10^{-2}$ \\ 
& $10^{-3}$ & 7.02$\times 10^{-1}$ & 8.87$\times 10^{-2}$ & 7.02$\times 10^{-1}$ & 9.73$\times 10^{-2}$ \\ 
& $10^{-2}$ & 6.77$\times 10^{-1}$ & 8.50$\times 10^{-2}$ & 6.76$\times 10^{-1}$ & 9.35$\times 10^{-2}$ \\ 
& $10^{-1}$ & 2.98$\times 10^{-1}$ & 4.10$\times 10^{-2}$ & 2.98$\times 10^{-1}$ & 4.10$\times 10^{-2}$ \\ 
\end{tabular}
\end{ruledtabular}
\label{SC-num-WKB-Z1-}
\end{table*}
                                                                              
\begin{table*}[tbp]
\caption{Fundamental frequencies for the
$Z_{2}^{-}$ field in RNdS, with $q=0.5$ and $m=1.0$, using the
numerical and semianalytical methods.}
\begin{ruledtabular}
\begin{tabular}{cccccc}

\multicolumn{2}{c}{q = 0.5} & \multicolumn{2}{c}{Numerical} &
\multicolumn{2}{c}{Semianalytical} \\ 
$\ell$ & $\Lambda$ & Re($\omega_0$) & -Im($\omega_0$) & Re($\omega_0$) & -Im($%
\omega_0$) \\ \hline
2 &$10^{-5}$ & 3.82$\times 10^{-1}$ & 8.96$\times 10^{-2}$ &
3.81$\times 10^{-1}$ & 8.98$\times 10^{-2}$ \\  
& $10^{-4}$ & 3.82$\times 10^{-1}$ & 8.96$\times 10^{-2}$ &
3.81$\times 10^{-1}$ & 8.98$\times 10^{-2}$ \\  
& $10^{-3}$ & 3.80$\times 10^{-1}$ & 8.93$\times 10^{-2}$ &
3.80$\times 10^{-1}$ & 8.95$\times 10^{-2}$ \\  
& $10^{-2}$ & 3.66$\times 10^{-1}$ & 8.67$\times 10^{-2}$ & 3.65$\times 10^{-1}$ & 8.67$\times 10^{-2}$ \\ 
& $10^{-1}$ & 1.60$\times 10^{-1}$ & 4.06$\times 10^{-2}$ & 1.60$\times 10^{-1}$ & 4.05$\times 10^{-2}$ \\ \hline
3 &$10^{-5}$ & 6.13$\times 10^{-1}$ & 8.59$\times 10^{-2}$ & 6.12$\times 10^{-1}$ & 9.33$\times 10^{-2}$ \\ 
& $10^{-4}$ & 6.13$\times 10^{-1}$ & 8.59$\times 10^{-2}$ & 6.12$\times 10^{-1}$ & 9.32$\times 10^{-2}$ \\ 
& $10^{-3}$ & 6.10$\times 10^{-1}$ & 8.59$\times 10^{-2}$ & 6.10$\times 10^{-1}$ & 9.29$\times 10^{-2}$ \\ 
& $10^{-2}$ & 5.88$\times 10^{-1}$ & 8.50$\times 10^{-2}$ & 5.87$\times 10^{-1}$ & 8.97$\times 10^{-2}$ \\ 
& $10^{-1}$ & 2.58$\times 10^{-1}$ & 4.07$\times 10^{-2}$ & 2.59$\times 10^{-1}$ & 4.07$\times 10^{-2}$ \\ 
\end{tabular}
\end{ruledtabular}
\label{SC-num-WKB-Z2-}
\end{table*}
                                                                              
\begin{table*}[tbp]
\caption{Fundamental frequencies for the
$Z_{1}^{+}$ field in RNdS, with
$q=0.5$ and $m=1.0$, using the numerical and semianalytical methods.}
\begin{ruledtabular}
\begin{tabular}{cccccc}

\multicolumn{2}{c}{q = 0.5} & \multicolumn{2}{c}{Numerical} &
\multicolumn{2}{c}{Semianalytical} \\ 
$\ell$ & $\Lambda$ & Re($\omega_0$) & -Im($\omega_0$) & Re($\omega_0$) & -Im($%
\omega_0$) \\ \hline
1 &$10^{-5}$ & 2.71$\times 10^{-1}$ & 9.51$\times 10^{-2}$ & 2.69$\times 10^{-1}$ & 9.55$\times 10^{-2}$ \\ 
& $10^{-4}$ & 2.71$\times 10^{-1}$ & 9.50$\times 10^{-2}$ & 2.68$\times 10^{-1}$ & 9.55$\times 10^{-2}$ \\ 
& $10^{-3}$ & 2.70$\times 10^{-1}$ & 9.47$\times 10^{-2}$ & 2.67$\times 10^{-1}$ & 9.51$\times 10^{-1}$ \\ 
& $10^{-2}$ & 2.60$\times 10^{-1}$ & 9.11$\times 10^{-2}$ & 2.57$\times 10^{-1}$ & 9.15$\times 10^{-2}$ \\ 
& $10^{-1}$ & 1.16$\times 10^{-1}$ & 4.08$\times 10^{-2}$ & 1.16$\times 10^{-1}$ & 4.08$\times 10^{-2}$ \\ \hline
2 &$10^{-5}$ & 4.94$\times 10^{-1}$ & 9.71$\times 10^{-2}$ & 4.93$\times 10^{-1}$ & 9.72$\times 10^{-2}$ \\ 
& $10^{-4}$ & 4.94$\times 10^{-1}$ & 9.71$\times 10^{-2}$ & 4.93$\times 10^{-1}$ & 9.72$\times 10^{-2}$ \\ 
& $10^{-3}$ & 4.92$\times 10^{-1}$ & 9.67$\times 10^{-2}$ & 4.91$\times 10^{-1}$ & 9.68$\times 10^{-2}$ \\ 
& $10^{-2}$ & 4.73$\times 10^{-1}$ & 9.30$\times 10^{-2}$ & 4.73$\times 10^{-1}$ & 9.31$\times 10^{-2}$ \\ 
& $10^{-1}$ & 2.09$\times 10^{-1}$ & 4.16$\times 10^{-2}$ & 2.09$\times 10^{-1}$ & 4.10$\times 10^{-2}$ \\ \hline
3 &$10^{-5}$ & 7.05$\times 10^{-1}$ & 8.95$\times 10^{-2}$ & 7.05$\times 10^{-1}$ & 9.77$\times 10^{-2}$ \\ 
& $10^{-4}$ & 7.05$\times 10^{-1}$ & 8.94$\times 10^{-2}$ & 7.05$\times 10^{-1}$ & 9.76$\times 10^{-2}$ \\ 
& $10^{-3}$ & 7.02$\times 10^{-1}$ & 8.86$\times 10^{-2}$ & 7.02$\times 10^{-1}$ & 9.73$\times 10^{-2}$ \\ 
& $10^{-2}$ & 6.77$\times 10^{-1}$ & 8.50$\times 10^{-2}$ & 6.76$\times 10^{-1}$ & 9.35$\times 10^{-2}$ \\ 
& $10^{-1}$ & 2.98$\times 10^{-1}$ & 4.10$\times 10^{-2}$ & 2.98$\times 10^{-1}$ & 4.10$\times 10^{-2}$ \\ 
\end{tabular}
\end{ruledtabular}
\label{SC-num-WKB-Z1+}
\end{table*}

\begin{table*}[tbp]
\caption{Fundamental frequencies for the
$Z_{2}^{+}$ field in RNdS, with $q=0.5$ and $m=1.0$, using the
numerical and semianalytical methods.}
\begin{ruledtabular}
\begin{tabular}{cccccc}

\multicolumn{2}{c}{q = 0.5} & \multicolumn{2}{c}{Numerical} &
\multicolumn{2}{c}{Semianalytical} \\ 
$\ell$ & $\Lambda$ & Re($\omega_0$) & -Im($\omega_0$) & Re($\omega_0$) & -Im($%
\omega_0$) \\ \hline
2 &$10^{-5}$ & 3.82$\times 10^{-1}$ & 8.96$\times 10^{-2}$ & 3.81$\times 10^{-1}$ & 8.99$\times 10^{-2}$ \\ 
& $10^{-4}$ & 3.82$\times 10^{-1}$ & 8.95$\times 10^{-2}$ & 3.81$\times 10^{-1}$ & 8.97$\times 10^{-2}$ \\ 
& $10^{-3}$ & 3.80$\times 10^{-1}$ & 8.93$\times 10^{-2}$ & 3.79$\times 10^{-1}$ & 8.94$\times 10^{-2}$ \\ 
& $10^{-2}$ & 3.66$\times 10^{-1}$ & 8.64$\times 10^{-2}$ & 3.65$\times 10^{-1}$ & 8.66$\times 10^{-2}$ \\ 
& $10^{-1}$ & 1.60$\times 10^{-1}$ & 4.13$\times 10^{-2}$ & 1.60$\times 10^{-1}$ & 4.05$\times 10^{-2}$ \\ \hline
3 &$10^{-5}$ & 6.13$\times 10^{-1}$ & 8.56$\times 10^{-2}$ & 6.12$\times 10^{-1}$ & 9.33$\times 10^{-2}$ \\ 
& $10^{-4}$ & 6.13$\times 10^{-1}$ & 8.56$\times 10^{-2}$ & 6.12$\times 10^{-1}$ & 9.32$\times 10^{-2}$ \\ 
& $10^{-3}$ & 6.10$\times 10^{-1}$ & 8.56$\times 10^{-2}$ & 6.10$\times 10^{-1}$ & 9.29$\times 10^{-2}$ \\ 
& $10^{-2}$ & 5.88$\times 10^{-1}$ & 8.49$\times 10^{-2}$ & 5.87$\times 10^{-1}$ & 8.97$\times 10^{-2}$ \\ 
& $10^{-1}$ & 2.58$\times 10^{-1}$ & 4.07$\times 10^{-2}$ & 2.58$\times 10^{-1}$ & 4.07$\times 10^{-2}$ \\ 
\end{tabular}
\end{ruledtabular}
\label{SC-num-WKB-Z2+}
\end{table*}

\begin{figure}
\resizebox{1\linewidth}{!}{\includegraphics*{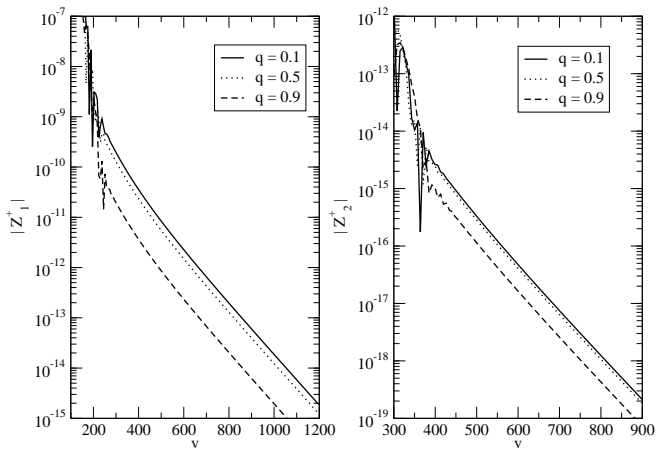}}
\caption{Tails of the
$Z_{1}^{+}$ ($\ell=1$) and $Z_{2}^{+}$ ($\ell=2$)  fields in RNdS. The
parameters for the geometry are $\Lambda=10^{-4}$ and $m=1.0$. The
results are similar for the other fields considered.}
\label{tails}
\end{figure}

For all fields considered, with a small enough
cosmological constant, there is a qualitative change in the behavior
of all fields considered. The late-time decay is dominated by an
exponential tail. Therefore in the RNdS geometry we have
\begin{eqnarray}
\psi_{\ell}^{sc}\sim e^{-k_{exp}^{sc}t} & \mathrm{with} & t\rightarrow\infty
\ ,
\end{eqnarray}
\begin{eqnarray}
Z_{1}^{\pm} \sim e^{-k_{exp}^{1\pm}t} & \mathrm{with} & t\rightarrow\infty \
,
\end{eqnarray}
\begin{eqnarray}
Z_{2}^{\pm} \sim e^{-k_{exp}^{2\pm}t} & \mathrm{with} & t\rightarrow\infty \
,
\end{eqnarray}
for $t$ sufficiently large. At the event and the cosmological horizons $t$
is substituted by $v$ and $u$, respectively. Figure \ref{tails}
illustrates this point, which was  noted in
\cite{Brady-97,Brady-99} for scalar fields, and we have extended this
consideration to coupled electromagnetic and gravitational
fields. In the aforementioned figure we compare the exponential tails
at the event horizon for exterior RNdS geometries.
                                                                              
An important point is that, unlike the quasinormal mode frequencies, the
exponential coefficients have
shown no dependence on the black hole's electric charge, for
all kinds of fields at hand. Close to  $\kappa_{c}=0$, our results are
compatible with the expressions
\begin{equation}
k_{exp}^{sc}(\kappa_{c})\approx\ell \left(\kappa_{c}+c^{sc}
\kappa_{c}^{2}\right) \ ,
\end{equation}
\begin{equation}
k_{exp}^{i \, \pm}(\kappa_{c}) \approx (\ell+1) \left(\kappa_{c}+c^{i \,
\pm}\kappa_{c}^{2}\right) \ \ \, i=1,2 \ ,
\end{equation}
for any $q$ lower than its extreme value.
The dynamics of the fields in de Sitter spacetimes is therefore very
different from similar cases in anti--de Sitter geometries, in which
for high charge there is an abrupt change in the fields' decay \cite{Wang-01}.
We have also explicitly assessed the behavior of the
$Z_{1,2}^{\pm }$ fields as $q\rightarrow 0$, comparing their quasinormal
frequencies and exponential tails to those observed in the SdS
$Z^{\pm }$ fields.  As anticipated, we found that the $Z_{2}^{+}$
field behaves like the SdS $Z^{+}$ field and that the $Z_{2}^{-}$ field
behaves like the SdS $Z^{-}$ field, if the charge is small enough. We have
observed that the RNdS fields
$Z_{2}^{\pm }$ tend smoothly to the SdS fields $Z^{\pm }$.

\section{Approaching the Asymptotically Flat Geometry}

Scalar fields in the SdS geometry near the asymptotically flat limit
were studied in \cite{Brady-97,Brady-99}. In this case there is
a clear separation between the event and the cosmological horizons,
such that 
\begin{equation}
\label{ggg}
\delta=\frac{r_{c}-r_{+}}{r_{+}} \gtrsim 50  \ .
\end{equation}

\begin{figure}
\setlength{\unitlength}{1.0mm}    
\resizebox{1\linewidth}{!}{\includegraphics*{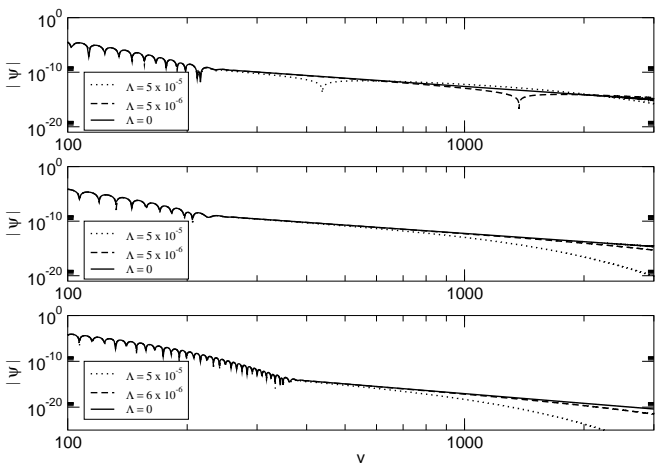}}
\caption{Approaching the asymptotically flat
geometry in SdS. Straight lines in the log-log graphs indicate
power law decay. In the graphs, $m=1.0$.}   
\label{appr-S}
\end{figure}

A new qualitative change occurs in this regime, namely, a decaying
phase with a power law behavior. Such a phase occurs between the
quasinormal mode decay and the exponential decay phases. The field
cannot be simply described by a superposition of the various modes,
which would imply a domination of the power law phase. This is
illustrated in Fig. \ref{appr-S}. 

\begin{figure}
\resizebox{1\linewidth}{!}{\includegraphics*{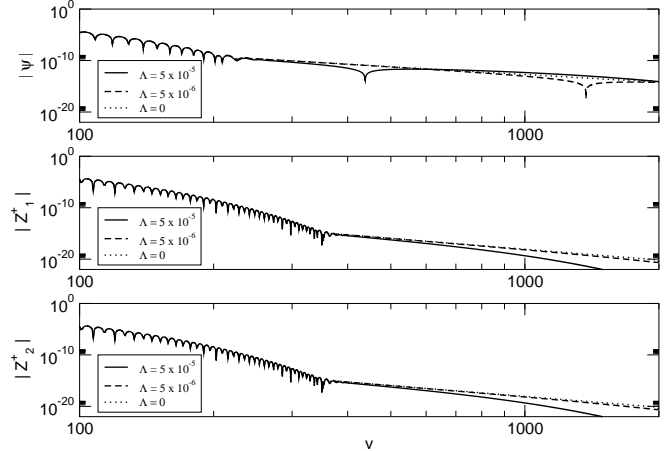}}
\caption{Scalar ($\ell=1$), $Z_{1}^{+}$ and
$Z_{2}^{+}$ fields ($\ell=2$), approaching the asymptotically flat
limit in RNdS. The parameters for the geometry are $q=0.5$ and $m=1.0$. The
results are similar for the other fields considered.}
\label{lim_RN}
\end{figure}

The situation for RNdS cases obeying Eq. (\ref{ggg}) is presented in
Fig.  \ref{lim_RN}. As can be seen in this figure, we
have a perfect power law tail developing for large $v$ when
$\Lambda=0$, as expected. For the RNdS exterior geometry with low
$\Lambda$ values, this power law tail appears quite clearly between
the quasinormal zone and the exponential tail. 

With such data, we can speak of three
different regimes in the field dynamics when one approaches the
asymptotically flat limit: first, a quasinormal regime, with its
characteristic damped oscillations, followed by an intermediate regime
for which the power law tail is visible, and a late-time region for
which an exponential tail dominates. This qualitative picture is valid
for all fields considered.

\section{Conclusions}

We have identified three regimes, according to the value of $\Lambda$ 
for the decay of the scalar, electromagnetic, and gravitational (or
$Z^\pm_{1,2}$ in RNdS) perturbations.  Near the extreme limit (high
$\Lambda$), we have analytic expressions for the effective potentials
and the quasinormal frequencies. The decay is entirely dominated by
the quasinormal modes (as in \cite{Beyer-99}),  that is, oscillatory
decay characterized by a nonvanishing real part of the quasinormal
frequency. 

In an intermediary parameter region (lower $\Lambda$), the wave
functions have an important qualitative change, with the appearance
of an exponential tail. This tail dominates the decay for large
time. Near the asymptotically flat limit ($\Lambda \ll 1$), we see an
intermediary phase between the quasinormal modes and the exponential
tail---a region of power law decay. When $\Lambda=0$, this region entirely
dominates the late-time behavior. 

Finally, for scalar fields with $\ell = 0$ a constant decay mode
appears, and its value depends on the $\dot{\psi}_{\ell}(0,x)$ initial
condition. Fig. \ref{non-char} reveals the appearance of the
constant value $\phi_0$  for large $t$, and its dependence on
$\dot{\psi}_{\ell}(0,x)$. The value of $\phi_0$ falls below  $10^{-7}$ for
$\dot{\psi}_{\ell}(0,x)=0$. These results are compatible with the
analytical predictions of \cite{Brady-99}. The analytical characterization of
these regions and the corresponding critical values of $\Lambda$ are crucial
to a better understanding of these qualitatively different regimes.

A very important aspect of the field dynamics in the RNdS geometries
is that the influence of the electric charge on the field behavior, in
general, was shown to be quite restricted. In particular, we
observed no dependence of the exponential tail coefficients with
$q$. This contrasts with previous results obtained in the anti--de
Sitter case \cite{Wang-01}, where the charge plays a fundamental role
in the tails. A deep understanding of this fact depends on
new analytical asymptotic results along the lines of the ones obtained in
\cite{Brady-99}. These points are now under investigation.
Also, since the presence of a charge implies an internal structure
similar to that of a rotating black hole, our results might be
interpreted as a broader universality of the frequencies here
obtained. This question deserves further study.


\begin{acknowledgments}
This work was supported by Funda\c{c}\~{a}o de Amparo
\`{a} Pesquisa do Estado de S\~{a}o Paulo (FAPESP), Conselho
Nacional de Desenvolvimento Cient\'{\i}fico e Tecnol\'{o}gico (CNPq)
and Coordena\c{c}\~{a}o de Aperfei\c{c}oamento de Pessoal de N\'{\i}vel
Superior (CAPES), Brazil. 
\end{acknowledgments}



\end{document}